\documentstyle[sprocl,epsfig]{article}

\def\be{\begin{equation}}
\def\ee{\end{equation}}
\def\bea{\begin{eqnarray}}
\def\eea{\end{eqnarray}}

\begin{document}

\title{BINDING ENERGY OF $\Lambda$ HYPERNUCLEI FROM REALISTIC $YN$
       INTERACTIONS}

\author{\underline{I. VIDA\~NA}, A. POLLS, A. RAMOS}

\address{Departament d'Estructura i Constituents de la Mat\`eria,
         Universitat de Barcelona, E-08028 Barcelona, Spain} 

\author{M. HJORTH-JENSEN}

\address{Nordita, Blegdamsvej 17, DK-2100 K\o benhavn \O, Denmark}


\maketitle\abstracts{ $s-$ and $p-$wave $\Lambda$ single--particle
energies are obtained for a variety of $\Lambda$ hypernuclei from the 
relevant self--energy constructed within the framework of a perturbative
many--body approach. Results are presented for present realistic 
hyperon--nucleon interactions such as J\"{u}lich B and Nijmegen SC 
models. Effects of the non--locality and energy--dependence of the 
self--energy on the bound states are investigated .}

\section{Introduction}
Several approaches have been followed to derive hyperon properties.
Traditionally, people have employed Woods-Saxon potentials which reproduce
quite well the measured
$\Lambda$ single--particle energies from medium to heavy \mbox{hypernuclei
\cite{motoba}}.  Non--localities and density dependent effects have been
included in non--relati--vistic Hartree--Fock
calculations with Skyrme $YN$ interactions in order to improve the overall fit to the
single--particle energies \cite{gal} . Hypernuclei have been studied
within a relativistic 
framework \cite{mares}, and microscopic
hypernuclear structure calculations have also been performed
\cite{halderson,kuo}.

Our work follows this last approach and its intention is to test the present $YN$ interactions
(i.e. J\"{u}lich and Nijmegen models). To this end we will evaluate the $s-$
and $p-$wave $\Lambda$
single--particle energies for some $\Lambda$ hypernuclei. Comparison with
experiment may help
in further constraining the $YN$ interactions.  
The starting point of this work is a nuclear matter $G-$matrix evaluated at a fixed nuclear density and
starting energy. This nuclear matter $G-$matrix is used to construct a finite
nucleus $G-$matrix which will
be employed in the evaluation of the hyperon self--energy. Finally, the real part of the hyperon self--energy
will be used as a non--local potential in a Schr\"{o}dinger equation in order
to get the single--particle energies
for the different orbits.  

\section{Formalism}
In this section we present a method to construct an effective $YN$ interaction in finite nuclei through an
expansion in terms of a nuclear matter $G-$matrix. The explicit details of
this method can be found in Ref. \cite{vidana}.  
The first step is to obtain the nuclear matter
$G-$matrix in terms of the bare interaction $V$ by solving the Bethe--Goldstone equation
\begin{equation}
G_{NM} = V + V \left( \frac{Q}{e} \right)_{NM} G_{NM} \ .
\end{equation}
It is important to note that this is a coupled channel problem because the hyperon of the intermediate 
state $YN$ can be a $\Lambda$ or a $\Sigma$.

Consider now the corresponding Bethe--Goldstone equation for the finite nucleus 
case
\begin{equation}
G_{FN} = V + V \left( \frac{Q}{e} \right)_{FN} G_{FN} \ .
\end{equation}
Eliminating the bare potential $V$ in both equations it is not difficult to
write $G_{FN}$ in terms
of $G_{NM}$ through an integral equation which can be truncated at second order
because the difference between the finite nucleus and the nuclear matter
intermediate propagators is small
\begin{equation}
G_{FN} \simeq G_{NM} + G_{NM} \left[ \left( \frac{Q}{e} \right)_{FN}
-\left( \frac{Q}{e} \right)_{NM} \right] G_{NM} \ .
\end{equation}
Finally, once we have $G_{FN}$, the hyperon self--energy at
Brueckner--Hartree--Fock level
reads in schematic form as
\begin{eqnarray}
\Sigma_{HF}=\sum_{N}\left\langle{Y}''N \right |G_{FN}\left |YN \right
\rangle 
\simeq \sum_{N}\left\langle{Y}''N \right |G_{NM}\left |YN \right \rangle \nonumber \\ 
+ \sum_{{Y}'N}\left\langle{Y}''N \right |G_{NM}\left |{Y}'N \right \rangle 
\left[ \left( \frac{Q}{e} \right)_{FN}-\left( \frac{Q}{e} \right)_{NM} \right]
\left\langle{Y}'N \right |G_{NM}\left |YN \right \rangle
\end{eqnarray}

\section{Results and Discussion}
In this section we show and discuss some of the results obtained for
$\Lambda$ hypernuclei. First, we have tested the stability of our
results against
variations of the nuclear density and the starting energy used in the calculation of the
nuclear matter $G-$matrix. 
We have found that the $1^{st}$ and $2^{nd}$ order terms, shown in eq. $(3)$, 
depend
quite strongly on these parameters which is an indication that the density
dependent effects
are important when the finite size of the nucleus is taken into account. Nevertheless, the 
whole calculation up to $2^{nd}$ order gives results very stable (see tables
I-IV of Ref. \cite{vidana}). 
In the next table we present the $s-$ and $p-$wave $\Lambda$ single--particle energies for a variety
of $\Lambda$ hypernuclei. All the results are given only for the J\"{u}lich B
interaction and no results are
shown for the Nijmegen SC because they appear clearly underbound with respect
the experimental data (e.g. in the case of $^{17}_{\Lambda}$O Nijmegen SC
gives -7.39 MeV in front of the experimental value -12.5 MeV).
\vspace{-0.25cm}
\begin{table}[h]
\caption{$\Lambda$ binding energies in the $1s_{1/2}$, $1p_{3/2}$ and
$1p_{1/2}$ orbits for different hypernuclei.}
\vspace{0.2cm}
\begin{center}
\footnotesize
\begin{tabular}{|c|c|c|c|c|}
\hline
\raisebox{0pt}[13pt][7pt]{Hypernuclei} &
\raisebox{0pt}[13pt][7pt]{Orbit} &\raisebox{0pt}[13pt][7pt]{$1^{st}$} &
\raisebox{0pt}[13pt][7pt]{$1^{st}+2p1h$} &
\raisebox{0pt}[13pt][7pt]{Exp.}   \\
\hline
\raisebox{0pt}[12pt][6pt]{$^{13}_{\Lambda}$C} &
\raisebox{0pt}[12pt][6pt]{$1s_{1/2}$} &
\raisebox{0pt}[12pt][6pt]{$-7.93$} &
\raisebox{0pt}[12pt][6pt]{$-9.48$} &
\raisebox{0pt}[12pt][6pt]{$-11.69$ ($^{13}_{\Lambda}$C) } \\
\hline
\raisebox{0pt}[12pt][6pt]{$^{17}_{\Lambda}$O} &
\raisebox{0pt}[12pt][6pt]{$1s_{1/2}$} &
\raisebox{0pt}[12pt][6pt]{$-10.15$} &
\raisebox{0pt}[12pt][6pt]{$-11.83$} &
\raisebox{0pt}[12pt][6pt]{$-12.5$ ($^{16}_{\Lambda}$O) } \\
{} & 
\raisebox{0pt}[12pt][6pt]{$1p_{3/2}$} &
{} &
\raisebox{0pt}[12pt][6pt]{$-0.87$} &
\raisebox{0pt}[12pt][6pt]{$-2.5$ ($1p$) } \\
{} &
\raisebox{0pt}[12pt][6pt]{$1p_{1/2}$} &
\raisebox{0pt}[12pt][6pt]{$-0.08$} &
\raisebox{0pt}[12pt][6pt]{$-1.06$} &
{} \\
\hline
\raisebox{0pt}[12pt][6pt]{$^{41}_{\Lambda}$Ca} &
\raisebox{0pt}[12pt][6pt]{$1s_{1/2}$} &
\raisebox{0pt}[12pt][6pt]{$-16.85$} &
\raisebox{0pt}[12pt][6pt]{$-19.60$} &
\raisebox{0pt}[12pt][6pt]{$-20$ ($^{40}_{\Lambda}$Ca) } \\
{} &
\raisebox{0pt}[12pt][6pt]{$1p_{3/2}$} &
\raisebox{0pt}[12pt][6pt]{$-6.70$} &
\raisebox{0pt}[12pt][6pt]{$-9.64$} &
\raisebox{0pt}[12pt][6pt]{$-12$ ($1p$) } \\
{} &
\raisebox{0pt}[12pt][6pt]{$1p_{1/2}$} &
\raisebox{0pt}[12pt][6pt]{$-6.92$} &
\raisebox{0pt}[12pt][6pt]{$-9.92$} &
{} \\
\hline
\raisebox{0pt}[12pt][6pt]{$^{91}_{\Lambda}$Zr} &
\raisebox{0pt}[12pt][6pt]{$1s_{1/2}$} &
\raisebox{0pt}[12pt][6pt]{$-22.24$} &
\raisebox{0pt}[12pt][6pt]{$-25.80$} &
\raisebox{0pt}[12pt][6pt]{$-23$ ($^{89}_{\Lambda}$Zr) } \\
{} &
\raisebox{0pt}[12pt][6pt]{$1p_{3/2}$} &
\raisebox{0pt}[12pt][6pt]{$-14.74$} &
\raisebox{0pt}[12pt][6pt]{$-18.19$} &
\raisebox{0pt}[12pt][6pt]{$-16$ ($1p$) } \\
{} &
\raisebox{0pt}[12pt][6pt]{$1p_{1/2}$} &
\raisebox{0pt}[12pt][6pt]{$-14.86$} &
\raisebox{0pt}[12pt][6pt]{$-18.30$} &
{} \\
\hline
\raisebox{0pt}[12pt][6pt]{$^{209}_{\Lambda}$Pb} &
\raisebox{0pt}[12pt][6pt]{$1s_{1/2}$} &
\raisebox{0pt}[12pt][6pt]{$-26.28$} &
\raisebox{0pt}[12pt][6pt]{$-31.36$} &
\raisebox{0pt}[12pt][6pt]{$-27$ ($^{208}_{\Lambda}$Pb) } \\
{} &
\raisebox{0pt}[12pt][6pt]{$1p_{3/2}$} &
\raisebox{0pt}[12pt][6pt]{$-21.22$} &
\raisebox{0pt}[12pt][6pt]{$-27.13$} &
\raisebox{0pt}[12pt][6pt]{$-22$ ($1p$) } \\
{} &
\raisebox{0pt}[12pt][6pt]{$1p_{1/2}$} &
\raisebox{0pt}[12pt][6pt]{$-21.30$} &
\raisebox{0pt}[12pt][6pt]{$-27.18$} &
{} \\
\hline
\end{tabular}
\end{center}
\end{table}

The agreement with experimental data is rather good, especially for the $s-$wave. Note that according to empirical
information the spin-orbit splitting is very small and note also that the $1p_{1/2}$ energy is lower than the $1p_{3/2}$
which is tied to the particular spin structure of the J\"{u}lich interaction.  

Finally, we show in the figure the $1s_{1/2}$ wave function of a $\Lambda$
in $^{17}_{\Lambda}$O obtained from our non--local self--energy (solid
line) or from a local Woods--Saxon potential (dashed line) of depth $-30.2$ MeV
and radious adjusted to reproduce the same binding energy. Nevertheless, as can
be seen our wave function is far more extended and this can have important
implications on the mesonic weak decay of hypernuclei. Only if we allow the
Woods--Saxon potential to have a shallower depth ($-23.6$ MeV) and a larger
radious we can not only reproduce the binding energy but also maximize the
overlap of the resulting wave function (dot--dashed line) with ours. 

\begin{figure}[h!]
   \caption{Wave function in r--space for the $1s_{1/2}$ $\Lambda$ in
$^{17}_{\Lambda}$O.}
       \setlength{\unitlength}{1mm}
       \begin{picture}(100,100)
       \put(17,35){\epsfxsize=8cm \epsfbox{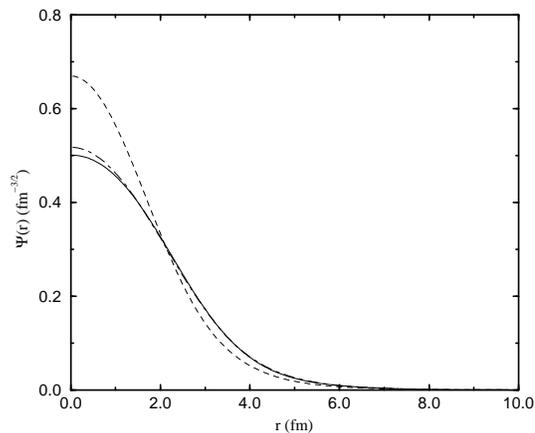}}
       \end{picture}
\end{figure}

\vspace*{-5.0cm}
\section{Conclusions}
We have developed a method to obtain an effective $YN$ interaction in finite
nuclei based on an expansion over a 
$G-$matrix calculated in nuclear matter at fixed nuclear density and starting energy. The truncation of this expansion
up to second order gives results very stable against variations of these parameters. $\Lambda$ single--particle energies
are well reproduced by the J\"{u}lich B model but appear clearly underbound by
the Nijmegen SC. Future perspectives of this
work are the study of scattering states and the study of the implications of our wave functions on the mesonic weak decay
of hypernuclei.

\end{document}